\documentclass[12pt,manuscript]{aastex}

\shorttitle{{\em Swift} and {\em RXTE} observations of GX 339-4 in the hard state}
\shortauthors{Allured, Tomsick, Kaaret, \& Yamaoka}

\usepackage{graphicx}
\usepackage{amsmath}
\usepackage{lscape}

\newcommand{\Msun}      {\mbox{$\rm\,M_{\mathord\odot}$}}

\begin{document}

\def\lsim{\mathrel{\lower .85ex\hbox{\rlap{$\sim$}\raise
.95ex\hbox{$<$} }}}
\def\gsim{\mathrel{\lower .80ex\hbox{\rlap{$\sim$}\raise
.90ex\hbox{$>$} }}}

\title{{\em Swift} and {\em RXTE} observations of the black hole transient GX 339-4 in the hard state between outbursts}

\author{Ryan Allured\altaffilmark{1}, John A. Tomsick\altaffilmark{2}, Philip Kaaret\altaffilmark{3}, and Kazutaka Yamaoka\altaffilmark{4}}

\altaffiltext{1}{602 Van Allen Hall, University of Iowa, IA 52242, USA
(e-mail: rallured@gmail.com)}
\altaffiltext{2}{Space Sciences Laboratory, 7 Gauss Way, 
University of California, Berkeley, CA 94720-7450, USA
(e-mail: jtomsick@ssl.berkeley.edu)}
\altaffiltext{3}{702 Van Allen Hall, University of Iowa, IA 52242, USA
(email: philip-kaaret@uiowa.edu)}
\altaffiltext{4}{Solar Terrestrial Environment Laboratory, Nagoya University,
 Furo-cho, Chikusa-ku, Nagoya 464-8601, Japan
(e-mail: yamaoka@stelab.nagoya-u.ac.jp)}

\begin{abstract}
 
We use simultaneous {\em Swift} and {\em RXTE} observations of the black hole binary GX 339-4 to measure the inner radius of its accretion disk in the hard state down to $0.4\%~L_{Edd}$ via modeling of the thermal disk emission and the relativistically broadened iron line.  For the luminosity range covered in this work, our results rule out a significantly truncated disk at 100--1000 $R_g$ as predicted by the advection-dominated accretion flow paradigm.  The measurements depend strongly on the assumed emission geometry, with most results providing no clear picture of radius evolution.  If the inclination is constrained to roughly 20$^\circ$, however, the measurements based on the thermal disk emission suggest a mildly receding disk at a luminosity of $0.4\%~L_{Edd}$.  The iron abundance varies between $\sim1-2$ solar abundances, with the $i=20^\circ$ results indicating a negative correlation with luminosity, though this is likely due to a change in disk illumination geometry.



\end{abstract}


\section{INTRODUCTION}

Stellar mass black holes are typically observed in binary orbits with a companion star.  Accretion of material from the companion star onto the black hole forms an accretion disk, which converts gravitational energy into radiation primarily in the X-ray band.  Accretion in black hole binaries (BHBs) can be a sporadic process, resulting in X-ray outbursts separated by periods of low luminosity.  As a BHB evolves over time, it can be observed to transition between three canonical emission states: thermal, hard, and steep power law \citep{RM06}.  The thermal state is generally characterized by high luminosity, a thermally dominated spectrum, and low levels of X-ray variability.  In the hard state, sources have spectra dominated by a power law component, relatively high X-ray variability, and are frequently accompanied by compact radio jets \citep{Fender06}.  While the hard state is typically seen at low levels of luminosity at the beginning or end of an outburst, a source can remain in the hard state even at high luminosities \citep{Homan01,Fender04,Tomsick05,Dunn10}.  The steep power law is not well understood, occurring at high luminosities with strong variability in the form of quasi-periodic oscillations \citep{vanderKlis06} and a steep power law index.  Note that BHBs are also frequently classified into states on the basis of model independent parameters such as spectral hardness and intensity (e.g.\ Belloni 2010).


The common framework for understanding BHB state transitions is described in \citet{Esin97}.  Within this model, BHBs are usually found in the hard state at low accretion rates ($\dot{m}\lesssim0.01 M_{Edd}$).  The inner part of the accretion disk is optically thin and is characterized by a low density, possibly caused by radiative evaporation \citep{Meyer00}.  Most of the viscously released energy is contained in the ions, which are unable to radiate efficiently in an advection-dominated accretion flow (ADAF).  This results in energy being advected toward the black hole rather than being radiated away.  Additionally, the flow becomes spherical due to the high pressures expected in the ion gas.  At these low accretion rates, the thermally-dominated accretion disk is predicted to be truncated at an inner radius on the order of 100--1000 $R_g$ ($R_g=GM/c^2$). At higher mass accretion rates (thermal state), the accretion disk reaches all the way down to the innermost stable circular orbit (ISCO).  The precise luminosity and/or mass accretion rate at which the truncation occurs, if at all, is still a matter of debate \citep{Miller06_2,Rykoff07,Tomsick08,Reis10}.

One way to test for the presence of an ADAF involves measuring the inner radius of the thermally-dominated accretion disk of BHBs at low luminosities.  The thermal emission of the disk is dependent on both its inner temperature and inner radius, and modeling the shape of the thermal component in low luminosity BHB spectra can provide measurements of both.  This is complicated, however, due to the power law domination of BHB spectra in the hard state; the low normalization and temperature of the thermal component are difficult to constrain.  There are also uncertainties in corrections required by spectral hardening via scattering in the corona and general relativistic boundary conditions at the inner radius \citep{Kubota98,Shimura95}.  Another method of radius measurement uses the shape of the relativistically broadened iron line \citep{Fabian89}.  The effect of relativistic broadening as a function of disk geometry has been calculated by \citet{Laor91}, but models based on these calculations fail to include thermal broadening inherent to the line.  A model has been incorporated into XSPEC that includes thermal broadening ({\tt reflionx}; \citet{Ross05}) and also calculates the emission lines and continuum due to reflection of emission off of the disk in a physically self-consistent manner.

GX 339-4 is a galactic BHB with a persistent iron line in its various emission states.  The distance has been estimated to be 6 kpc $<D<$ 15 kpc by \citet{Hynes04} via analysis of the Na and Ca lines in the optical spectrum of the companion star.  \citet{Hynes03} calculated the mass function to be 5.8 $M_\odot$ using radial velocity data of the companion star.  There is debate in the literature regarding the inclination, with some results at $\sim 20-30^\circ$ \citep{Miller06,Done10}, and a more recent study favoring $\sim 50^\circ$ \citep{Shidatsu11}.  There is also debate on the inner accretion disk radius of GX 339-4 at low luminosities.  \citet{Tomsick09} have measured the radius at the lowest luminosity (0.14\% $L_{Edd}$) and find a significantly truncated disk at $>65~R_g$.  At slightly higher luminosities ($\sim2-4\%~L_{Edd}$), however, several results indicate a disk radius consistent with the ISCO \citep{Miller06,Reis08}, while \citet{Shidatsu11} find a marginally truncated disk at $13.3^{+6.0}_{-6.4}~R_g$.  This paper examines the inner disk radius of GX 339-4 using both thermal modeling and relativistic line broadening at a variety of luminosities in the hard state.

\section{OBSERVATIONS AND DATA REDUCTION}

\begin{figure}
\begin{center}
\includegraphics[width=6 in]{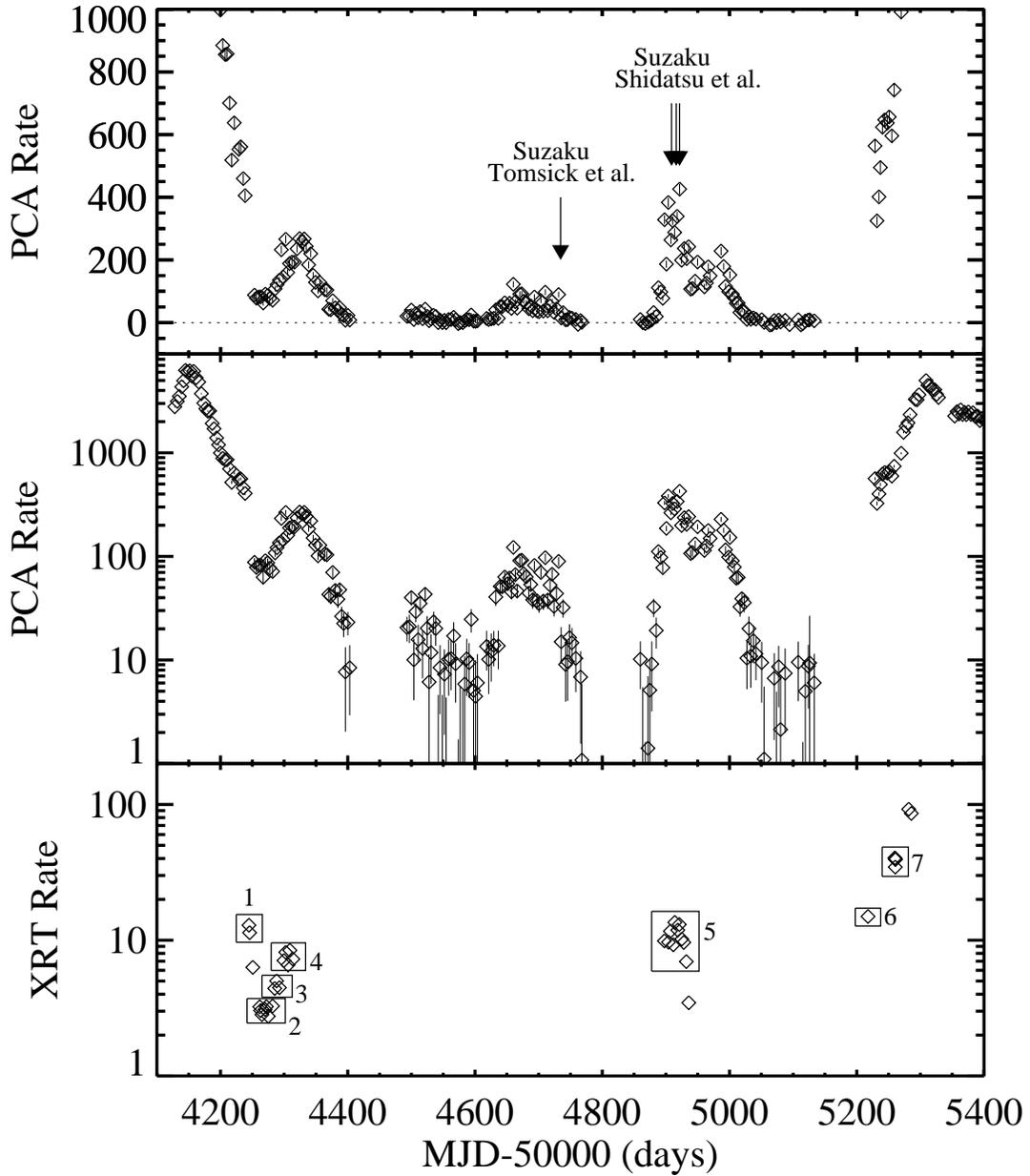}
\caption{The PCA X-ray count rate taken from the {\em RXTE} Bulge Scans website is shown on both a linear (top) and logarithmic (middle) scale from late 2006 to mid 2010.  The energy range covers 2--10 keV from 5 PCUs.  The XRT X-ray count rate is shown over the same time period on the bottom, with boxes indicating the grouping of the seven spectra in this paper.  Spectra 1 and 2 contain observations from \citet{Tomsick08}.  The times at which the observations from \citet{Tomsick09} and \citet{Shidatsu11} were taken are indicated in the top panel.\label{fig:lc}}
\end{center}
\end{figure}

We analyzed {\em Swift} X-ray Telescope (XRT; \citet{Burrows05}) and {\em RXTE} Proportional Counter Array (PCA; \citet{Jahoda06}) observations of GX 339-4 between its 2007 and 2010 outbursts.  The observations span 2007 May 24 to 2010 March 6 (MJD 54244--55261), with total exposures of 52,628 s and 114,160 s for the XRT and PCA, respectively.  The data was reduced using HEASOFT v6.11.  PCA spectra were extracted using {\tt saextract}, with {\tt marfrmf} 3.2.6 used to produce the response matrices and the SkyVLE model to produce the background spectra.  XRT spectra were extracted using {\tt xselect}, with {\tt swxwt0to2s6\_20010101v013.rmf} used as the response and {\tt xrtmkarf\_0.5.9} to produce the ancillary response file. Due to the proximity of GX 339-4 to the Galactic plane, observations from March 2001, February 2002, and September 2003 were used to create a quiescent spectrum that was subtracted from all PCA spectra.  To improve statistics, data were combined into seven PCA/XRT spectra as indicated by Table\ \ref{tab:obs}.  The PCA and XRT lightcurves are shown in Fig.\ \ref{fig:lc} with boxes indicating spectral groupings.  Spectral fitting was performed in XSPEC v12.7.1.  XRT spectra were restricted to 0.5--10 keV and PCA spectra to 3--25 keV.  Systematic errors of 0.5\% were added to the PCA spectra.

\begin{table}
\caption{Times, exposures, and count rates for {\em Swift} and {\em RXTE} observations\label{tab:obs}}
\begin{minipage}{\linewidth}
\footnotesize
\begin{tabular}{ccccccc} \hline \hline
Observation \# & MJD start & MJD end & XRT Exposure & XRT Rate (c/s) & PCA Exposure & PCA Rate (c/s)\\ \hline
1 & 54,244.7 & 54,246.0 &  7,119~s (2 obs)  & 11.54 &  7,984~s (3 obs)  & 60.79\\
2 & 54,261.7 & 54,282.0 & 14,830~s (7 obs)  &  3.06 & 49,952~s (24 obs) & 20.89\\
3 & 54,285.2 & 54,292.6 &  3,854~s (3 obs)  &  4.61 &  9,344~s (8 obs)  & 37.00\\
4 & 54,299.7 & 54,313.9 &  4,985~s (5 obs)  &  7.36 & 12,800~s (8 obs)  & 57.25\\
5 & 54,897.6 & 54,932.2 & 14,184~s (10 obs) & 10.39 & 14,784~s (6 obs)  & 84.90\\
6 & 55,217.7 & 55,218.0 &  4,756~s (1 obs)  & 14.97 &  2,800~s (1 obs)  & 102.6\\
7 & 55,260.0 & 55,261.1 &  2,900~s (4 obs)  & 39.72 & 16,496~s (5 obs)  & 252.6\\ \hline\hline
\end{tabular}
\end{minipage}
\end{table}

\section{RESULTS}

\subsection{Spectral Model and Fitting Procedure}
\label{sec:Procedure}

\begin{figure}
\begin{center}
\includegraphics[width=6 in]{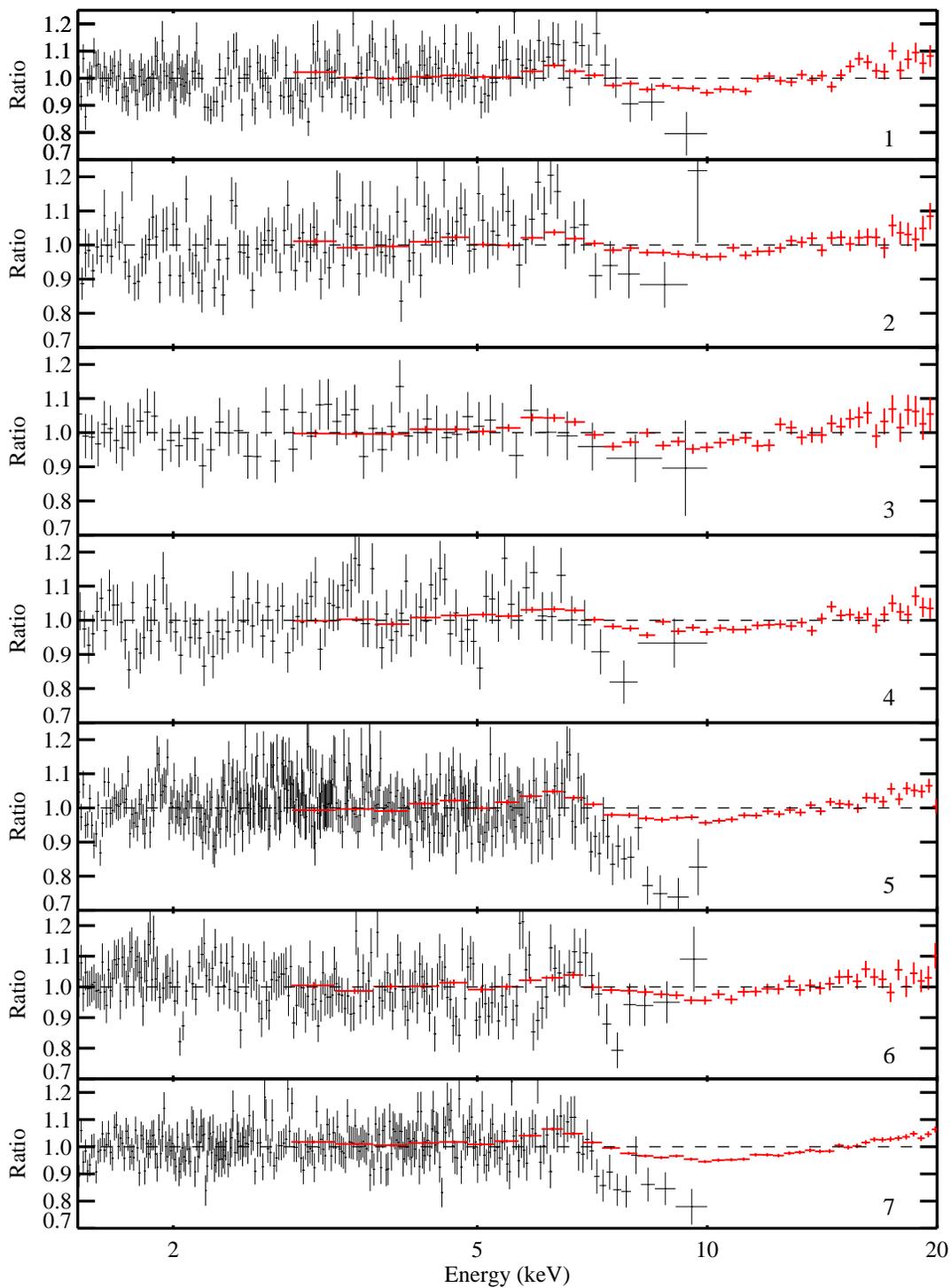}
\caption{Data/Model ratios for both PCA (red) and XRT (black) data.  A broad iron line is evident in all spectra, as well as a reflection continuum above 10 keV.\label{fig:ratio}}
\end{center}
\end{figure}

To investigate the possible presence of an iron line, we examined the residuals in the 5--8 keV range.  Specifically, all seven spectra were fit with an absorbed multicolor blackbody disk model plus a power law component, the standard model for BHBs \citep{RM06}.  The data/model ratios in Fig.\ \ref{fig:ratio} show a clear, broad peak near 6-7 keV in the PCA data.  The peak is also evident in the XRT residuals, except in spectra 3 and 4.  The PCA residuals above 10 keV show a positive deviation from the power law, a feature typical of disk reflection.  Confident in the presence of a broad iron line in all seven spectra, we switched to a model including iron line emission with relativistic blurring and disk reflection.


PCA and XRT spectra were fit simultaneously using an absorbed ({\tt phabs}) multicolor blackbody disk model ({\tt diskbb}; \citet{Mitsuda84}) plus a standard power law component.  The \citet{Wilms00} interstellar medium abundances and \citet{BCMC} photoelectric cross sections were used in the {\tt phabs} component.  Spectral features due to reflection of the power law component off of the disk were modeled using {\tt reflionx} \citep{Ross05}.  {\tt reflionx} assumes a folding energy of 300 keV in the power law, which was included with the multiplicative component {\tt highecut}.  The power law and reflection model components were convolved with {\tt kdblur} (a relativistic broadening model using the calculations from \citet{Laor91}) in order to account for relativistic effects from the compact object.

To allow for a difference in PCA and XRT calibrations, we applied a normalization constant to the PCA model and left it free to vary during the fits.  All {\tt diskbb} and {\tt powerlaw} parameters were left free.  The physical {\tt reflionx} parameters of ionization parameter and and iron abundance were left free, but the photon index was fixed to that of the {\tt powerlaw} component for consistency.  The redshift was set to 0 due to the galactic nature of the source, and the normalization was left variable.  The {\tt kdblur} inner radius parameter was left free, the outer radius was assumed to be 400 $R_g$ ($R_g = GM/c^2$), and the disk inclination and emissivity index were varied between separate fitting rounds.  As discussed in the introduction, there is a discrepancy between prior results on the inclination $i$ of the binary system.  Additionally, the emissivity index $e$, where line emission scales with disk radius as $R^{-e}$, is poorly constrained due to unknown accretion geometry.  Recent results indicate varying emissivity indices in black hole accretion disks \citep{Svoboda12,Wilkins12}, but {\tt kdblur} does not incorporate a varying emissivity index.  We carried out a total of six rounds of fitting using $i=20^\circ,50^\circ$ and $e=2.1,3.0,4.0$.  The lower $e$ value of 2.1 was chosen to avoid flux divergence at 2.0, and the upper $e$ value of 4.0 was chosen to investigate values higher than the Newtonian value of 3.0.

\begin{figure}[h]
\begin{center}
\includegraphics[scale=.5,angle=270]{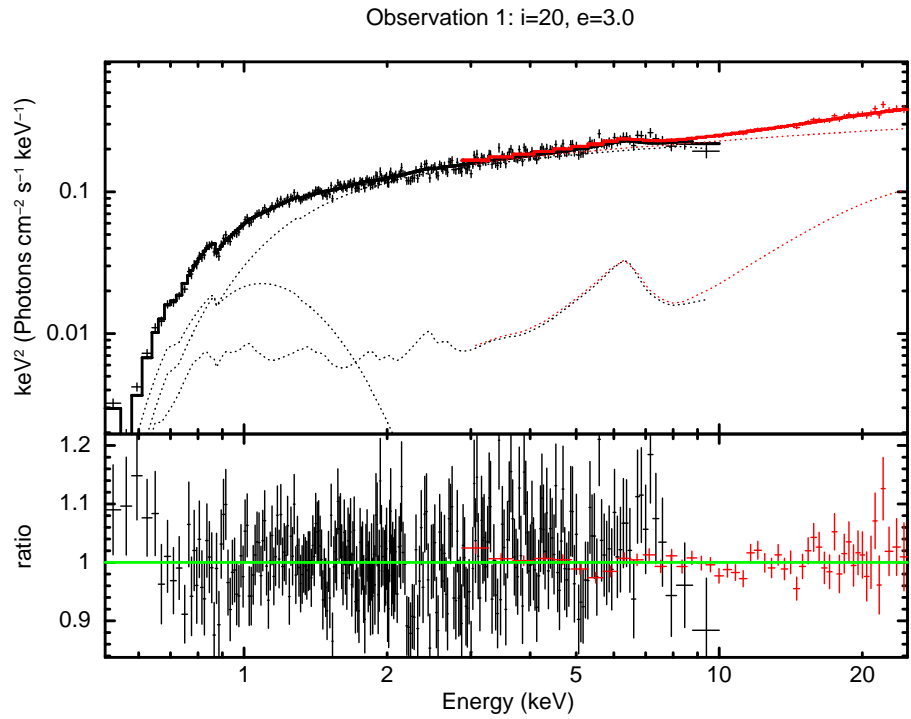}
\vspace{0.0cm}
\caption{The energy spectrum and data:model ratios for observation 1.  PCA data are shown in red, and XRT data are shown in black.  The disk inclination is fixed to 20$^\circ$ and the emissivity index is fixed to 3.0.\label{fig:i20Spectrum}}
\end{center}
\end{figure}

\begin{figure}[h]
\begin{center}
\includegraphics[scale=.5,angle=270]{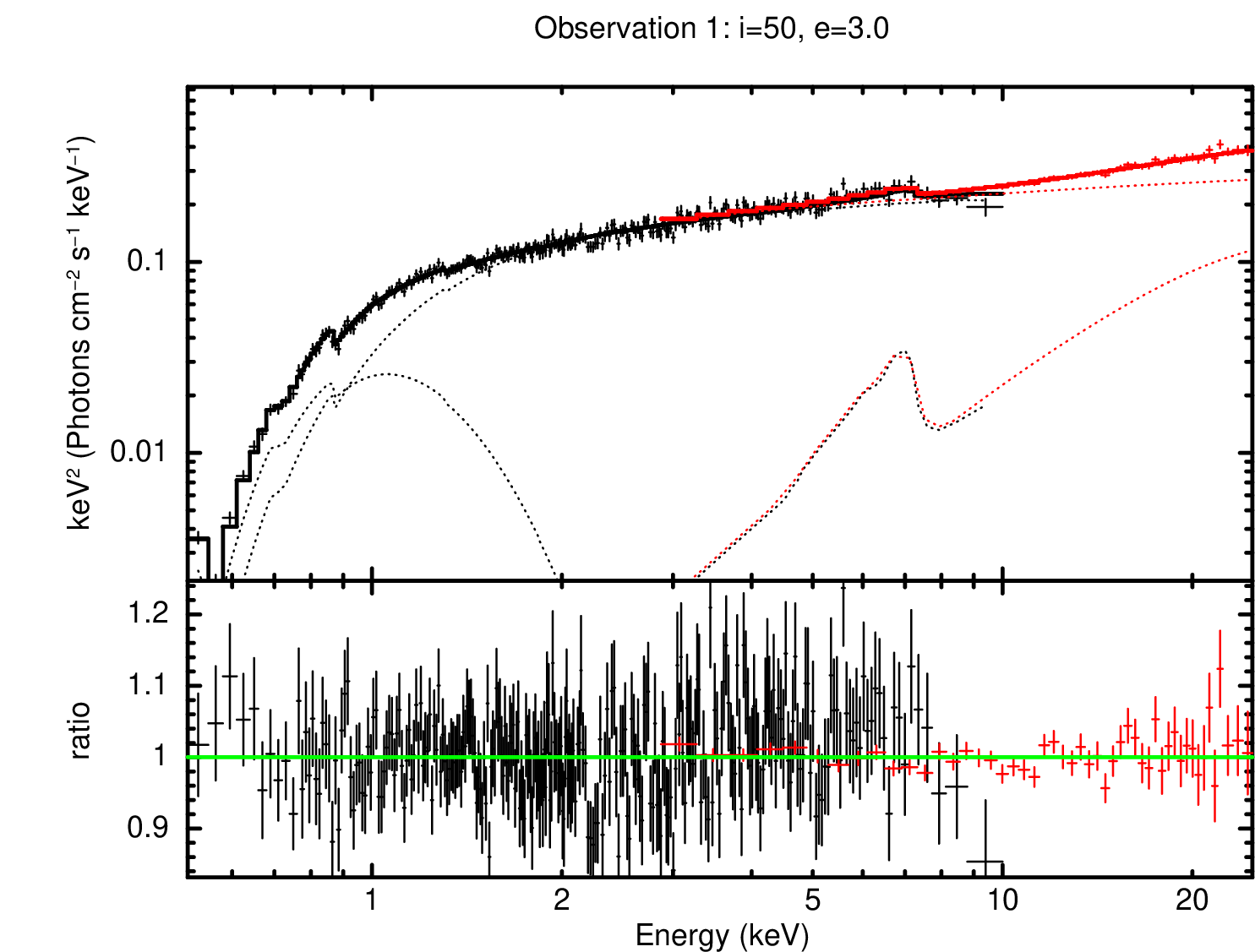}
\vspace{0.0cm}
\caption{The energy spectrum and data:model ratios for observation 1.  PCA data are shown in red, and XRT data are shown in black.  The disk inclination is fixed to 50$^\circ$ and the emissivity index is fixed to 3.0.\label{fig:i50Spectrum}}
\end{center}
\end{figure}

\citet{Miller09} used {\em Chandra} High Energy Transmission Grating spectra to show that individual absorption edge column densities remained constant despite state transitions in several BHBs, motivating a fixed hydrogen column density.  Thus, all fits were repeated after the hydrogen column density was fixed to a best fit constant of 7.91$\times10^{21}$ cm$^{-2}$ based on initial fits.  Additionally, the {\tt diskbb} component was not required in any of the fits to observation 6.  Characteristic energy spectra for observation 1 are shown in Figs.\ \ref{fig:i20Spectrum} ($i=20^\circ$,$e=3.0$) and \ref{fig:i50Spectrum} ($i=50^\circ$,$e=3.0$).  Spectral parameters of all seven observations for the $i=20^\circ$,$e=3.0$ fitting round are shown in Table \ref{tab:ryanfit}.  The rows give, in descending order, the inner disk temperature, {\tt diskbb} normalization, {\tt kdblur} inner radius, photon index, power law normalization, iron abundance, {\tt reflionx} ionization parameter, {\tt reflionx} normalization, best fit $\chi^2$, and model luminosity.

\begin{landscape}
\begin{table}
\caption{$i=20^\circ$,$e=3.0$ Fitting Round Spectral Parameters\label{tab:ryanfit}}
\begin{minipage}{\linewidth}
\footnotesize
\begin{tabular}{cccccccc}
\hline \hline
& Obs 1 & Obs 2 & Obs 3 & Obs 4 & Obs 5 & Obs 6 & Obs 7 \\ \hline
$kT_{\rm in}$ (keV) & $0.189^{+0.007}_{-0.008}$ & $0.129^{+0.020}_{-0.017}$ & $0.161^{+0.026}_{-0.035}$ & $0.170^{+0.025}_{-0.014}$ & $0.129^{+0.019}_{-0.016}$ & NA & $0.258^{+0.027}_{-0.025}$\\
$N_{\rm DBB}$/1,000 & $24^{+8}_{-5}$ & $34^{+56}_{-24}$ & $13^{+31}_{-11}$ & $19^{+16}_{-12}$ & $83^{+132}_{-51}$ & NA & $4^{+3}_{-2}$\\
$R_{in}$ ($GM/c^2$) & $6.7^{+9.1}_{-2.3}$ & $3.6^{+1.9}_{-0.9}$ & $12.8^{+19.8}_{-9.0}$ & $6.9^{+9.1}_{-3.7}$ & $19.5^{+25.0}_{-8.5}$ & $16.3^{+11.7}_{-5.2}$ & $19.7^{+12.1}_{-6.5}$ \\
$\Gamma$ & $1.73^{+0.02}_{-0.01}$ & $1.60\pm0.01$ & $1.61\pm0.02$ & $1.60\pm0.01$ & $1.59\pm0.01$ & $1.61\pm0.02$ & $1.65\pm0.01$\\
$N_\Gamma$ (keV$^{-1}$cm$^{-2}$s$^{-1}$) & $0.130\pm0.004$ & $0.035\pm0.001$ & $0.083^{+0.003}_{-0.002}$ & $0.090\pm0.002$ & $0.118\pm0.002$ & $0.161\pm0.003$ & $0.422\pm0.005$\\
Fe abundance (solar) & $2.00^{+0.51}_{-0.47}$ & $2.00^{+1.37}_{-0.22}$ & $2.40^{+1.47}_{-0.62}$ & $2.00^{+1.89}_{-0.25}$ & $1.03^{+0.25}_{-0.08}$ & $0.95^{+0.32}_{-0.17}$ & $1.65^{+0.24}_{-0.23}$\\
$\xi$ (erg cm/s) & $210^{+11}_{-6}$ & $268^{+144}_{-44}$ & $200^{+36}_{-74}$ & $211^{+45}_{-34}$ & $209^{+7}_{-6}$ & $202^{+9}_{-42}$ & $212^{+5}_{-3}$\\
$N_{\rm R}\times10^6$ & $10.9\pm1.8$ & $1.9^{+0.7}_{-1.1}$ & $10.7^{+4.9}_{-3.0}$ & $8.6^{+3.7}_{-2.1}$ & $15.1^{+1.4}_{-1.2}$ & $23.5^{+6.2}_{-3.0}$ & $49.8^{+3.2}_{-3.1}$\\ \hline
$\chi^2/\nu$ & 360.21/323 & 297.98/228 & 98.91/122 & 214.16/186 & 628.4/459 & 404.3/312 & 475.82/388\\
$L_{1-100}/L_{Edd}$, 10 \Msun, 8 kpc & 1.18\% & 0.42\% & 1.08\% & 1.12\% & 1.68\% & 2.23\% & 5.21\%\\
\hline
\end{tabular}
\end{minipage}
\end{table}
\end{landscape}

\subsection{Inner Radius Determination}
\label{sec:RadDet}

Inner radii were determined from the {\tt kdblur} $R_{in}$ parameter and by using the radial dependence of the {\tt diskbb} normalization parameter.  The normalization parameter is defined as $\sqrt{N_{\rm DBB}/{\rm cos}i}=r_{in}({\rm km})/d(10~{\rm kpc})$, where $i$ is the inclination of the disk, $r_{in}$ is the inner radius, and $d$ is the distance to the object.  Physical radii obtained from $N_{\rm DBB}$ must be converted into units of gravitational radii.  We assume a distance of 8 kpc for convenient comparison to prior studies. The inclination was taken to be either $20^\circ$ or $50^\circ$ depending on the fitting round.  In order to derive a physical inner radius, correction factors due to general relativistic boundary conditions and spectral hardening must be taken into account.  Based on \citet{Kubota98} and \citet{Shimura95}, we write the physical radius as $R_{in}=\xi \kappa^2 r_{in}$, and have adopted a relativistic correction factor $\xi=0.412$ and a spectral hardening factor $\kappa =1.7$.  Finally, we convert $R_{in}$ to units of gravitational radii $R_{g}=GM/c^2$.  We take the black hole mass to be 10 M$_\odot$, again for the purposes of convenient comparison.

\begin{figure}[h]
\begin{center}
\vspace{-1cm}
\includegraphics[width=.9\linewidth]{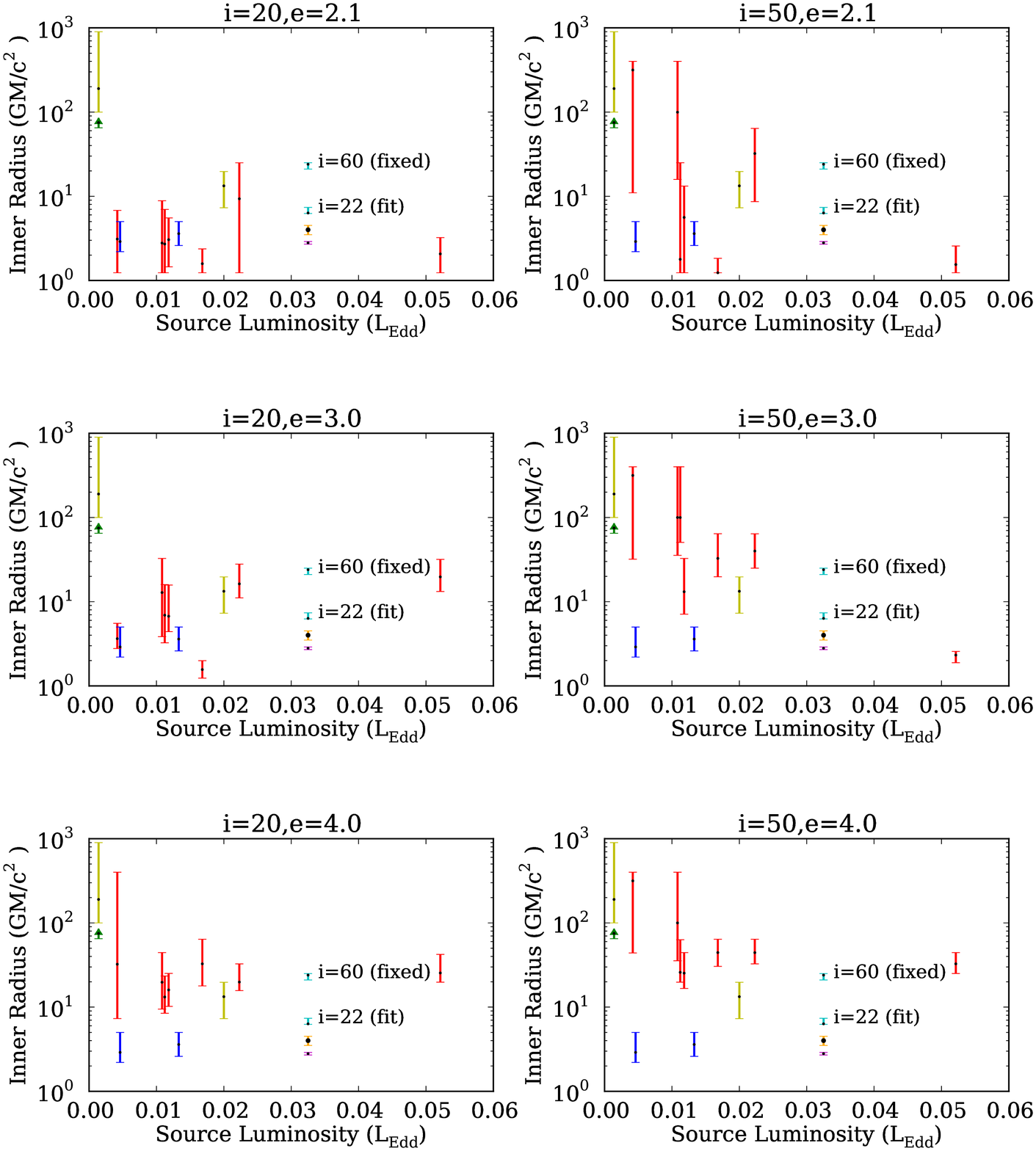}
\vspace{-1cm}
\caption{Inner disk radius obtained from {\tt kdblur} versus source luminosity are shown in red.  Results from all six fitting rounds are shown along with relevant results from Table 6 of \citet{Shidatsu11}.  Blue points are from \citet{Tomsick08}, yellow points are from \citet{Shidatsu11}, teal points are from \citet{Done10}, the green lower limit is from \citet{Tomsick09}, the orange point is from \citet{Miller06}, and the purple point is from \citet{Reis08}.\label{fig:KdblurRins}}
\end{center}
\end{figure}

\begin{figure}[h]
\begin{center}
\vspace{-2cm}
\includegraphics[width=.9\linewidth]{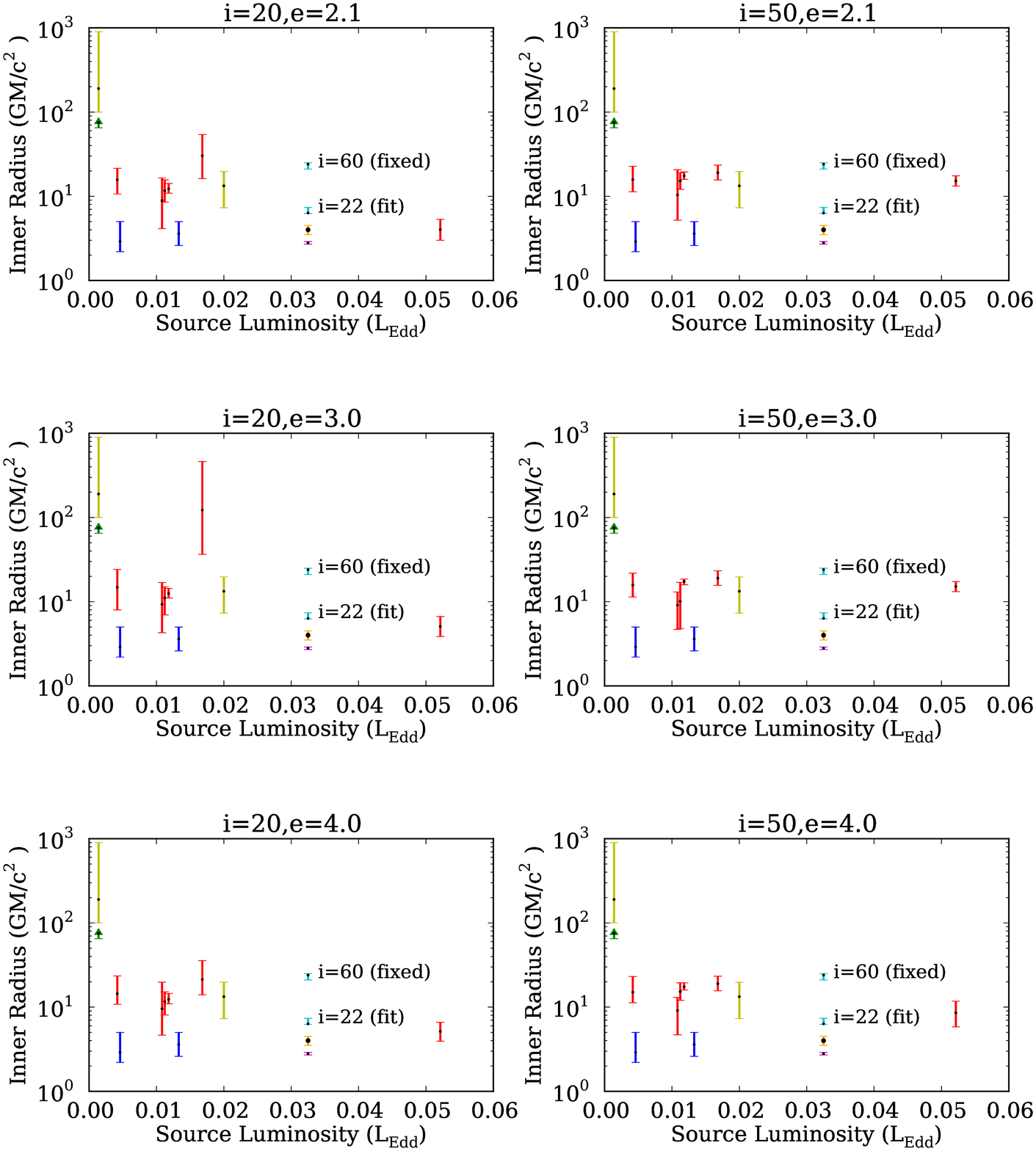}
\vspace{-1cm}
\caption{Inner disk radii obtained from {\tt diskbb} are shown in red.  Observation 6 is not shown due to the lack of a significant disk component.  The other points are as in Fig.\ \ref{fig:KdblurRins}.\label{fig:DnormRins}}
\end{center}
\end{figure}

To investigate the evolution of disk radius with luminosity, we calculate luminosities for each observation in units of Eddington luminosity.  The Eddington luminosity is estimated using a black hole mass of 10 M$_\odot$, and our source luminosity is estimated using the unabsorbed model flux from 1--100 keV and a distance of 8 kpc.  Inner radii obtained from {\tt kdblur} are plotted against source luminosity in Fig.\ \ref{fig:KdblurRins} along with the relevant results from Table 6 of \citet{Shidatsu11}, and those obtained from {\tt diskbb} are shown in Fig.\ \ref{fig:DnormRins}.  Error bars are 90\% confidence limits.  All of these results used the same distance and mass assumptions as above to determine luminosity in units of $L_{Edd}$.

\subsection{Correlation Analysis}

\begin{figure}[h]
\begin{center}
\vspace{-2cm}
\includegraphics[width=.9\linewidth]{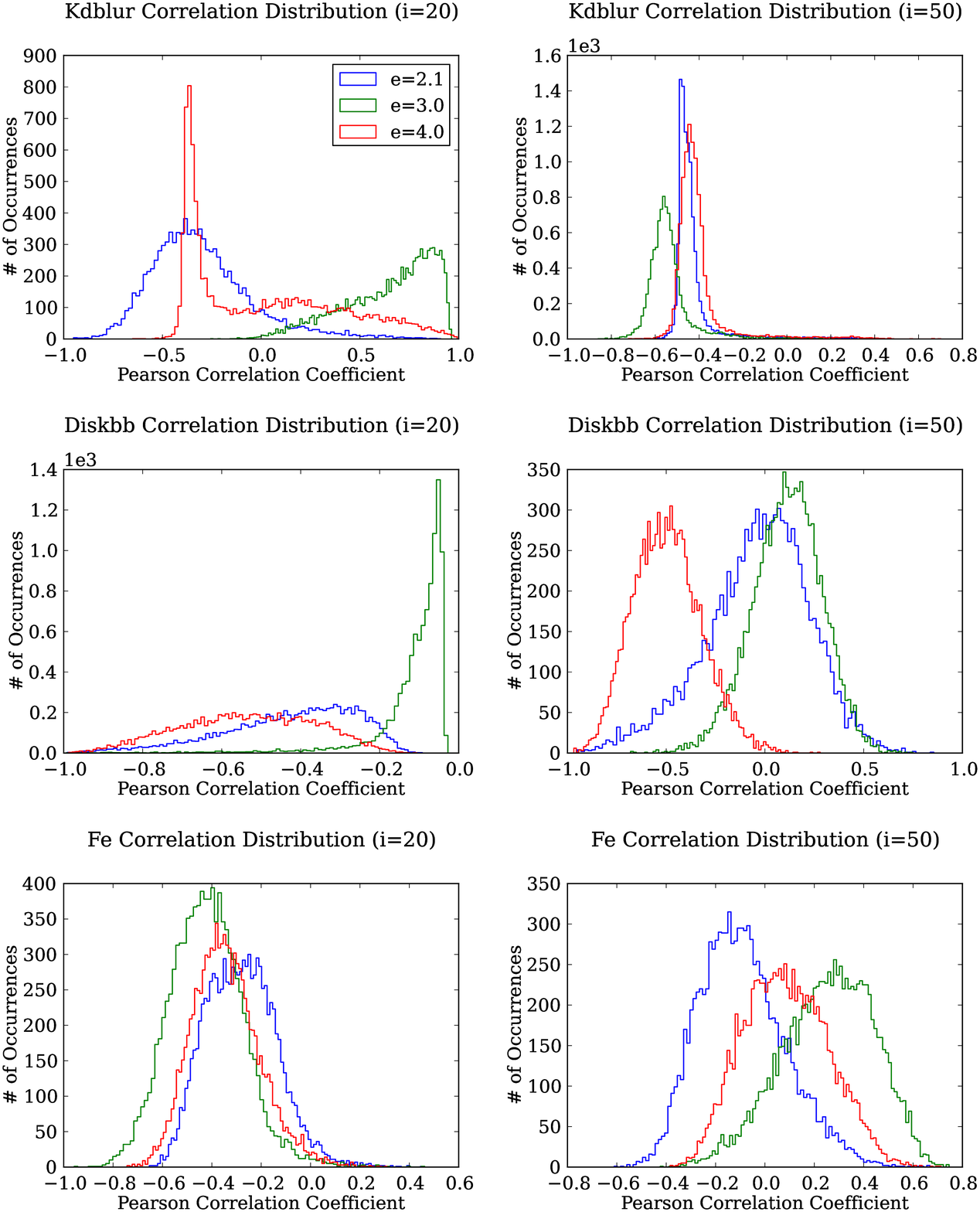}
\vspace{-1cm}
\caption{Monte-Carlo simulated Pearson luminosity correlation coefficient distributions for {\tt kdblur} $R_{in}$, inner radii obtained from {\tt diskbb} normalization, and the iron abundance for $i=20^\circ$ and $i=50^\circ$ fitting rounds.\label{fig:CorPlots}}
\end{center}
\end{figure}

Two statistical tests were performed on the inner radii obtained from each fitting round in order to quantify the relationship the inner disk radius has with luminosity.  First, a one-tailed test was used to determine whether the radii were consistent with a constant.  The data were fit to a constant, and the resulting $\chi^2$ value was compared to the $\chi^2$ distribution with the appropriate degrees of freedom (5 for {\tt diskbb}, 6 for {\tt kdblur}).  A $\chi^2$ value in the upper tail of the distribution led to a rejection of the null hypothesis that the radius was constant.  Second, the Pearson correlation coefficient distribution was simulated to determine how the radius depended on luminosity.  For each observation and fitting round, $10^5$ inner radii were generated based on the asymmetric 90\% confidence regions and assuming Gaussian statistics.  From this Monte-Carlo generated data, $10^5$ Pearson correlation coefficients were calculated and binned into a histogram.  Because it was noticed that the iron abundance has a large variation in all of the fitting rounds, these tests were also carried out on the iron abundance parameter.  The P-value for rejection of the null hypotheses and the average correlation coefficients ($\rho$) with standard deviations for each fitting round are given in Table \ref{tab:CorResults}.  To be clear, a low P-value is cause to reject the hypothesis of a constant parameter.  Plots of the correlation coefficient distributions are shown in Fig.\ \ref{fig:CorPlots}.

\begin{table}
\caption{Luminosity Correlation Analysis Results\label{tab:CorResults}}
\begin{minipage}{\linewidth}
\footnotesize
\begin{tabular}{c|cc|cc|cc} \hline \hline
Fitting Round & {\tt kdblur} $R_{in}~\rho$ & P-Value & {\tt diskbb} $R_{in}~\rho$ & P-Value &  Fe $\rho$ & P-Value \\ \hline
$i=20,e=2.1$ & $-0.304\pm0.258$ & 0.428 & $-0.425\pm0.169$ & 6.23e-13 & $-0.281\pm0.143$ & 3.10e-6\\
$i=20,e=3.0$ & $0.645\pm0.228$ & 1.41e-11 & $-0.113\pm0.101$ & 2.94e-7 & $-0.422\pm0.150$ & 3.78e-4\\
$i=20,e=4.0$ & $0.011\pm0.381$ & 0.401 & $-0.536\pm0.169$ & 2.87e-9 & $-0.343\pm0.137$ & 6.72e-8\\
$i=50,e=2.1$ & $-0.429\pm0.109$ & 0.0313 & $-0.038\pm0.264$ & 0.484 & $-0.092\pm0.175$ & 8.00e-5\\
$i=50,e=3.0$ & $-0.537\pm0.105$ & 1.01e-9 & $0.112\pm0.175$ & 9.47e-3 & $0.255\pm0.187$ & 1.42e-8\\
$i=50,e=4.0$ & $-0.404\pm0.130$ & 0.203 & $-0.495\pm0.173$ & 4.55e-9 & $0.084\pm0.170$ & 7.26e-9\\ \hline\hline
\end{tabular}
\end{minipage}
\end{table}

\section{DISCUSSION}

The inner radii obtained from the {\tt kdblur} component depend strongly on the assumed disk geometry.  It is clear from Fig.\ \ref{fig:KdblurRins} that there is a general lack of correlation of the inner radii from our various fitting rounds.  We attribute this to insufficient energy resolution and/or count rates to reliably distinguish between Compton/thermal broadening in {\tt reflionx} and relativistic broadening in {\tt kdblur}.  Because of this, it is difficult to draw conclusions about the inner radius evolution based on the {\tt kdblur} $R_{in}$ parameter without independent constraints on the inner disk geometry.  While the Pearson correlation coefficient distributions based on {\tt kdblur} $R_{in}$ shown in Fig.\ \ref{fig:CorPlots} for $i=50^\circ$ (top left) show strong negative correlation between inner radius and luminosity, suggestive of the beginnings of disk recession, only the $i=50^\circ,e=3.0$ fitting round has a P-value inconsistent with a constant radius (see Table \ref{tab:CorResults}).  The distributions obtained for $i=20^\circ$ (middle left) are clearly ambiguous, and only the $i=20^\circ,e=3.0$ fitting round has a P-value inconsistent with a constant radius.

The {\tt kdblur} component does have a minor effect on modeling of the thermal disk emission, as can be seen from the energy spectra in Figs.\ \ref{fig:i20Spectrum} and \ref{fig:i50Spectrum}.  At $i=20^\circ$, the reduced Doppler broadening from {\tt kdblur} is compensated by thermal broadening, resulting in a higher {\tt reflionx} ionization parameter.  A higher ionization parameter increases the overall strength of the reflection component and necessitates a smaller {\tt reflionx} normalization.  The stronger reflection at soft energies then suppresses the {\tt diskbb} component by lowering its normalization, though in observation 5 the inner temperature is driven low instead.


The dependence of inner radii obtained from {\tt diskbb} on assumed disk geometry is much weaker.  Our results shown in Fig.\ \ref{fig:DnormRins} show a similar functional form of inner radius between the fitting rounds.  Note, however, the higher radius of observation 7 at $i=50^\circ$ due to the effect of {\tt kdblur} described above.  Because this observation occurs at the highest luminosity (5.21\% $L_{Edd}$), it has a high impact on the correlation distributions and causes the $i=50^\circ,e=2.1$ and $i=50^\circ,e=3.0$ fitting rounds to be consistent with a constant radius.  The other four fitting rounds, all three $i=20^\circ$ rounds in particular, have P-values highly inconsistent with a constant radius and are negatively correlated at $\gtrsim 99.9\%$ confidence (none of the 1000 simulated $\rho$ were positive for these three cases).  Thus, our results at $i=20^\circ$ based on {\tt diskbb} would constitute strong evidence of a mildly receding inner disk with independent inclination constraints.


The P-values in Table \ref{tab:CorResults} all indicate a varying iron abundance.  The strength of an iron emission line depends on the amount of iron in the source, but also on how that iron is illuminated (i.e. strength and incidence angle of the illuminating flux).  Relativistic effects at the inner disk could produce a wide range of illumination geometries, including a varying emissivity index as pointed out in \S \ref{sec:Procedure}.  Indeed, \citet{Bhayani10} have investigated variability of relativistic iron lines in Seyfert galaxies using rms spectra, and frequently find enhanced variability with respect to the continuum in the red wings of iron lines that are thought to originate at the inner disk.  The {\tt reflionx} model does not account for a varying emissivity index, meaning our varying iron abundance could actually be indicating a change in this geometry.  The iron abundances obtained in the $i=20^\circ$ fits are negatively correlated with luminosity, a phenomenon discovered in Seyferts by \citet{Nandra97}.  This monotonic relationship of iron abundance is consistent with the change in disk geometry suggested by our $i=20^\circ$ {\tt diskbb} inner radii.  \citet{Nandra97} suggest a more ionized disk at higher luminosities to explain this relationship, though our {\tt reflionx} ionization parameter shows no signs of positive correlation with luminosity.  At $i=50^\circ$, the relationship of iron abundance with luminosity is weakly and ambiguously correlated.

If the inner disk inclination of GX 339-4 could be independently constrained to $\sim 20^\circ$, our results would imply a truncated disk at lower luminosities.  However, none of our results are consistent with an inner radius at the hundreds of gravitational radii suggested by the ADAF model.  Previous results at $0.14\%~L_{Edd}$ clearly indicate a significantly truncated disk, so we may have evidence of the beginning of the transition to an ADAF.  In this case, a more dramatic dependence of inner radius on luminosity is expected at $\lesssim 0.4\%~L_{Edd}$.  More observations between $0.1-1.0\%~L_{Edd}$ are required to fully understand the inner radius evolution.  If relativistic broadening is used to measure the inner radius, thermal broadening must be taken into account and the energy resolution of the observations must be good enough to separate the two broadening mechanisms.  If thermal disk modeling is used, deep observations with energy coverage well below 3 keV are required to constrain the shape of the soft spectral component.  In either case, the observations must have excellent statistics in order to provide robust measurements.

\acknowledgments

The authors thank the anonymous referee for thoughtful criticisms which have improved this work.  Ryan Allured and Philip Kaaret acknowledge partial support from NASA grant NNX08AY58G.  John A.\ Tomsick acknowledges partial support from NASA Swift Guest Investigator grants NNX10AF94G and NNX10AK36G.

\bibliographystyle{jwapjbib}
\bibliography{refs}

\end{document}